\documentclass[pra,preprint,tightenlines]{revtex4}

\usepackage{amsmath}
\usepackage{amssymb}
\usepackage{longtable}
\usepackage{caption}
\usepackage{dcolumn}

\newcolumntype{b}{D{(}{\ (}{-1}}
\renewcommand{\textsc}[1]{$\,${\footnotesize #1}}

\newcommand{\Tref}[1]{Table~\ref{#1}}

\newcommand{\cm}{cm$^{-1}$}
\newcommand{\comment}[1]{}
\newcommand{\smallcite}[1]{\cite{#1}}

\begin{document}

\title{Laboratory spectroscopy and the search for space-time variation of the fine structure constant using QSO spectra.}

\author{J. C. Berengut}
\email{jcb@phys.unsw.edu.au}
\affiliation{School of Physics, University of New South Wales, Sydney 2052, Australia}
\author{V. A. Dzuba}
\affiliation{School of Physics, University of New South Wales, Sydney 2052, Australia}
\author{V. V. Flambaum}
\email{flambaum@phys.unsw.edu.au}
\affiliation{School of Physics, University of New South Wales, Sydney 2052, Australia}
\author{M. G. Kozlov}
\affiliation{Petersburg Nuclear Physics Institute, Gatchina, 188300, Russia}
\author{M. V. Marchenko}
\affiliation{School of Physics, University of New South Wales, Sydney 2052, Australia}
\author{M. T. Murphy}
\affiliation{Institute of Astronomy, University of Cambridge, Madingley Road, Cambridge CB3 0HA, UK}
\author{J. K. Webb}
\affiliation{School of Physics, University of New South Wales, Sydney 2052, Australia}

\date{1 March 2006}

\begin{abstract}

Theories unifying gravity with other interactions suggest spatial and temporal variation of fundamental ``constants'' in the Universe. A change in the fine structure constant, $\alpha = e^2/\hbar c$, could be detected via shifts in the frequencies of atomic transitions in quasar absorption systems. Previous studies of three independent samples of data, containing 143 absorption systems spread from 2 to 10~billion years after big bang, hint that $\alpha$ was smaller 7 -- 11~billion years ago \cite{webb99prl,webb01prl,murphy01mnrasA,murphy01mnrasC,webb03ass,murphy03mnras}.

To continue this study we urgently need accurate laboratory measurements of atomic transition frequencies. The aim of this paper is to provide a compilation of transitions of importance to the search for $\alpha$ variation. They are $E1$ transitions to the ground state in several different atoms and ions, with wavelengths ranging from around 900 -- 6000~\AA, and require an accuracy of better than 10$^{-4}$~\AA. We also discuss isotope shift measurements that are needed in order to resolve systematic effects in the study.

Researchers who are interested in performing these measurements should contact the authors directly. 

\end{abstract}


\maketitle

Current theories that seek to unify gravity with the other fundamental interactions suggest the possibility of spatial and temporal variation of fundamental constants in the Universe (see, for example the review~\cite{uzan03rmp}). Several studies have tried to probe the values of constants at earlier stages in the evolution of the universe. One method compares atomic transition frequencies on Earth and in quasar (QSO) absorption spectra, and recent studies of these spectra have revealed hints that the fine structure constant, $\alpha$, was smaller in the early universe \cite{webb99prl,webb01prl,murphy01mnrasA,murphy01mnrasC,webb03ass,murphy03mnras}. The latest results of this group, which combine data from 143 absorption systems over the redshift range $0.2 < z_\textrm{\small abs} < 3.7$, show $\Delta \alpha / \alpha = (-0.543 \pm 0.116) \times 10^{-5}$ \cite{murphy03mnras}. However, attempts to replicate this result using a similar analysis, but different data sets from different telescopes, indicate no variation of $\alpha$ \cite{quast04aap,srianand04prl,chand04aap,levshakov05aap}.

To continue this work and resolve the discrepancies, several new transitions are being considered. In \Tref{tab:wavelengths} we present a list of lines commonly observed in high-resolution QSO spectra. Those transitions for which high-accuracy laboratory data are needed are marked with either `A' (very important) or `B' (mildly important). Some of the lines marked `A' have already been measured quite accurately (references are given), but even in these cases confirmation and improvement is still urgently required. Predominantly, the wavelengths given in \Tref{tab:wavelengths} come from the compilation \cite{morton91ajss,morton03ajss} and have errors of about 0.005~\AA, although it is possible that some errors are closer to 0.05~\AA. Note that the oscillator strengths presented are not as accurate as the wavelengths: these measurements are much more difficult. As a general rule, the lines are more important for $\alpha$ variation if they lie above 1215.67~\AA\ (the Lyman-$\alpha$ line of hydrogen) due to the ``Lyman-$\alpha$ forest'' seen in QSO spectra.

Isotope shift measurements for these transitions are also needed in order to resolve a source of systematic error in the variation of $\alpha$ studies: the isotope abundance ratios in the gas clouds sampled in the quasar absorption spectra may not match those on Earth \cite{murphy01mnrasB,murphy03ass}. Accurate measurements of the isotope shift are required to quantify these systematic effects. They can also be used to determine the abundances in the early universe directly, to test models of nuclear processes in stars. In addition to the transitions previously mentioned, in \Tref{tab:wavelengths}, we present lines that were used in previous studies (and have precise wavelength measurements), but for which the isotopic structure has not been measured. These transitions are marked with an `I'.

We previously calculated the relativistic energy shifts, or $q$ values \cite{dzuba02praA,berengut04praB,berengut05pra,berengut06pra}. We have presented them here for easy reference. The difference between the transition frequencies in QSO spectra ($\omega$) and in the laboratory ($\omega_0$) depends on the relative values of $\alpha$. The dependence of the frequencies on small changes in $\alpha$ is given by the formula $\omega~=~\omega_0~+~qx$, where $x~=~(\alpha/\alpha_0)^2~-~1$. The $q$ values are calculated using atomic physics codes. The atomic energy levels are calculated to a first approximation using relativistic Hartree-Fock (Dirac-Hartree-Fock). Higher order effects are taken into account using many-body perturbation theory for single-valence-electron systems, or using configuration interaction for many-valence-electron systems. Both methods assume a frozen Hartree-Fock core. The value of $\alpha$ is varied in the computer codes and the energy levels are recalculated, and hence the transition frequencies. The $q$ values are extracted as
\[
	q = \frac{d \omega}{d x} \bigg|_{x=0}
\]
We also account for complications due to level pseudo-crossing \cite{dzuba02praA}.

\section*{Acknowledgments}

The authors would like to thank Don~Morton, Scott~Bergeson, and Wim~Ubachs for useful comments and for pointing out some errors.

\newpage

\begin{longtable}{lcccbcc}
\caption[List of high-priority lines observed in QSO spectra.]
{\label{tab:wavelengths} High-priority lines observed in QSO spectra. The first column is the ion; the second and third columns are the rest wavelength and the transition frequency, respectively; the fourth column is the oscillator strength; the relativistic shift ($q$ value) is shown in the fifth column, where known. Those transitions for which high-accuracy laboratory data are needed are marked with either `A' (very important) or `B' (mildly important). Lines for which there is no measured isotopic structure are marked `I'. Additionally, there are some lines for which both the transition frequency and isotope shift are known; these are marked with an `M'. They are included here for reference only (of course, verification would still be useful). References for precisely measured lines are given in the last column. The second set of references are for isotope shift measurements, where available. The wavelengths and oscillator strengths are taken from the compilation by Morton \cite{morton91ajss,morton03ajss}.}\\
\hline \hline
Atom/ & Wavelength     &  Frequency      & Oscillator & \multicolumn{1}{c}{$q$ value} & \comment{Status} & Refs. \\
Ion   & $\lambda$ (\AA)& $\omega_0$ (\cm)& Strength   & \multicolumn{1}{c}{(\cm)}     &  & \\
\hline \hline \endfirsthead
\caption[]{(continued)} \\
\hline \hline
Atom/ & Wavelength     &  Frequency      & Oscillator & \multicolumn{1}{c}{$q$ value} &  & Refs. \\
Ion   & $\lambda$ (\AA)& $\omega_0$ (\cm)& Strength   & \multicolumn{1}{c}{(\cm)}     &  & \\
\hline \hline \endhead
\hline \endfoot
C\textsc{I}
    &  945.188 & 105799.1 & 0.272600 & 130(60) & M & \smallcite{labazan05pra} \\*
    & 1139.793 & 87735.30 & 0.013960 &  ~0(100)& B \\*
    & 1155.809 & 86519.47 & 0.017250 & " & B \\*
    & 1157.186 & 86416.55 & 0.549500 & " & B \\*
    & 1157.910 & 86362.52 & 0.021780 & " & B \\*
    & 1188.833 & 84116.09 & 0.016760 & " & B \\*
    & 1193.031 & 83820.13 & 0.044470 & " & B \\*
    & 1193.996 & 83752.41 & 0.009407 & " & B \\*
    & 1260.736 & 79318.78 & 0.039370 & 30(10) & A \\*
    & 1276.483 & 78340.28 & 0.004502 & 17(10) & A \\*
    & 1277.245 & 78293.49 & 0.096650 & -13(10)& A \\*
    & 1280.135 & 78116.74 & 0.024320 & -21(10)& A \\*
    & 1328.833 & 75253.97 & 0.058040 & 117(10)& A \\*
    & 1560.309 & 64089.85 & 0.080410 & 137(10)& A \\*
    & 1656.928 & 60352.63 & 0.140500 & -24(10)& A \\
C\textsc{II}
    & 1037.018 & 96430.32 & 0.123000 & 168(10) & A \\*
    & 1335.662 & 74869.20 & 0.012770 & 178(10) & A \\*
    & 1335.707 & 74866.68 & 0.114900 & 181(10) & A \\
C\textsc{III}
    &  977.020 & 102352.0 & 0.762000 & 165(10) & B \\
C\textsc{IV}
    & 1548.204 & 64590.99 & 0.190800 & 232(20) & A & \smallcite{griesmann00apj} \\*
    & 1550.781 & 64483.65 & 0.095220 & 104(20) & A & \smallcite{griesmann00apj} \\
O\textsc{I}
    & 1025.762 & 97488.54 & 0.020300 & ~0(20) & B \\*
    & 1026.476 & 97420.72 & 0.002460 & " & B \\*
    & 1039.230 & 96225.05 & 0.009197 & " & B \\*
    & 1302.168 & 76794.98 & 0.048870 & " & A \\
Na\textsc{I} 
    & 3303.320 & 30272.58 & 0.013400 & 59(4) & B \\*
    & 3303.930 & 30266.99 & 0.006700 & 53(4) & B \\*
    & 5891.583 & 16973.37 & 0.655000 & 63(4) & M & \smallcite{juncar81met}\smallcite{pescht77zpa,huber78prc} \\*
    & 5897.558 & 16956.17 & 0.327000 & 45(4) & M & \smallcite{juncar81met}\smallcite{gangrsky98epja} \\
Mg\textsc{I}
    & 2026.477 & 49346.73 & 0.112000 & 87 & I & \smallcite{aldenius06mnras} \\*
    & 2852.963 & 35051.27 & 0.181000 & 86(10) & M & \smallcite{aldenius06mnras,pickering98mnras}\smallcite{hallstadius79zpa,boiteux88jpf} \\
Mg\textsc{II}
    & 1239.925 & 80650.04 & 0.000267 & & B \\*
    & 2796.354 & 35760.85 & 0.612300 & 211(10) & M & \smallcite{aldenius06mnras,pickering98mnras}\smallcite{drullinger80ap} \\*
    & 2803.532 & 35669.30 & 0.305400 & 120(10) & I & \smallcite{aldenius06mnras,pickering98mnras} \\
Al\textsc{II}
    & 1670.789 & 59851.97 & 1.880000 & 270(30) & M & \smallcite{griesmann00apj} \\
Al\textsc{III}
    & 1854.718 & 53916.54 & 0.539000 & 464(30) & M & \smallcite{griesmann00apj} \\*
    & 1862.791 & 53682.88 & 0.268000 & 464(30) & M & \smallcite{griesmann00apj} \\
Si\textsc{II}
    & 1190.416 & 84004.26 & 0.250200 & & B \\*
    & 1193.290 & 83801.95 & 0.499100 & & B \\*
    & 1260.422 & 79338.50 & 1.007000 & & A \\*
    & 1304.370 & 76665.35 & 0.094000 & & A \\*
    & 1526.707 & 65500.45 & 0.117094 &  50 (30) & I & \smallcite{griesmann00apj} \\*
    & 1808.013 & 55309.34 & 0.002010 & 520 (30) & I & \smallcite{griesmann00apj} \\*
Si\textsc{IV}
    & 1393.760 & 71748.64 & 0.528000 & 862 & A & \smallcite{griesmann00apj} \\*
    & 1402.773 & 71287.54 & 0.262000 & 346 & A & \smallcite{griesmann00apj} \\
S\textsc{II}
    & 1250.583 & 79962.61 & 0.005350 & & A \\*
    & 1253.808 & 79756.83 & 0.010700 & & A \\*
    & 1259.518 & 79395.39 & 0.015900 & & A \\
Ca\textsc{II}
    & 3934.775 & 25414.41 & 0.688000 & 452 & A & \smallcite{morton03ajss} \\*
    & 3969.590 & 25191.52 & 0.341000 & 224 & A & \smallcite{morton03ajss} \\
Ti\textsc{II}
    & 1910.600 & 52339.58 & 0.202000 & -1564(150) & A \\*
    & 1910.938 & 52330.32 & 0.098000 & -1783(300) & A \\*
    & 3067.245 & 32602.55 & 0.041500 & 791(50) & I & \smallcite{aldenius06mnras} \\*
    & 3073.877 & 32532.21 & 0.104000 & 677(50) & I & \smallcite{aldenius06mnras} \\*
    & 3230.131 & 30958.50 & 0.057300 & 673(50) & I & \smallcite{aldenius06mnras} \\*
    & 3242.929 & 30836.32 & 0.183000 & 541(50) & I & \smallcite{aldenius06mnras} \\*
    & 3384.740 & 29544.37 & 0.282000 & 396(50) & I & \smallcite{aldenius06mnras} \\
Cr\textsc{II}
    & 2056.256 & 48632.06 & 0.105000 & -1110 (150) & I & \smallcite{aldenius06mnras,pickering00mnras} \\*
    & 2062.236 & 48491.05 & 0.078000 & -1280 (150) & I & \smallcite{aldenius06mnras,pickering00mnras} \\*
    & 2066.164 & 48398.87 & 0.051500 & -1360 (150) & I & \smallcite{aldenius06mnras,pickering00mnras} \\
Mn\textsc{II}
    & 1197.184 & 83529.35 & 0.156600 & -2556(450) & B \\*
    & 1199.391 & 83375.65 & 0.105900 & -2825(450) & B \\*
    & 1201.118 & 83255.77 & 0.088090 & -3033(450) & B \\*
    & 2576.877 & 38806.66 & 0.288000 & 1420(150) & I & \smallcite{aldenius06mnras} \\*
    & 2594.499 & 38543.08 & 0.223000 & 1148(150) & I & \smallcite{aldenius06mnras} \\*
    & 2606.462 & 38366.18 & 0.158000 &  986(150) & I & \smallcite{aldenius06mnras} \\
Fe\textsc{II}
    & 1063.176 & 94057.80 & 0.060000 & & B \\*
    & 1063.971 & 93987.52 & 0.003718 & & B \\*
    & 1096.877 & 91167.92 & 0.032400 & & B \\*
    & 1121.975 & 89128.55 & 0.020200 & & B \\*
    & 1125.448 & 88853.51 & 0.016000 & & B \\*
    & 1143.226 & 87471.77 & 0.017700 & & B \\*
    & 1144.939 & 87340.98 & 0.106000 & & B \\*
    & 1260.533 & 79331.52 & 0.025000 & & A \\*
    & 1608.450 & 62171.63 & 0.058000 & -1300 (300) & A & \smallcite{pickering02aap} \\*
    & 1611.200 & 62065.53 & 0.001360 & 1100 (300) & A & \smallcite{pickering02aap} \\*
    & 2249.877 & 44446.88 & 0.001821 & & A \\*
    & 2260.780 & 44232.51 & 0.002440 & & I  & \smallcite{aldenius06mnras}\\*
    & 2344.212 & 42658.24 & 0.114000 & 1210 (150) & I & \smallcite{aldenius06mnras,nave91josab} \\*
    & 2367.589 & 42237.06 & 0.000212 & 1904 & A \\
    & 2374.460 & 42114.83 & 0.031300 & 1590 (150) & I & \smallcite{aldenius06mnras,nave91josab} \\*
    & 2382.764 & 41968.06 & 0.320000 & 1460 (150) & I & \smallcite{aldenius06mnras,nave91josab} \\*
    & 2586.649 & 38660.05 & 0.069180 & 1490 (150) & I & \smallcite{aldenius06mnras,nave91josab} \\*
    & 2600.172 & 38458.99 & 0.238780 & 1330 (150) & I & \smallcite{aldenius06mnras,nave91josab} \\
Ni\textsc{II}
    & 1317.217 & 75917.64 & 0.146000 & & A \\*
    & 1370.132 & 72985.67 & 0.076900 & & A \\*
    & 1393.324 & 71770.82 & 0.022220 & & A \\*
    & 1454.842 & 68735.99 & 0.032300 & & A \\*
    & 1467.259 & 68154.29 & 0.009900 & & A \\*
    & 1467.756 & 68131.22 & 0.006300 & & A \\*
    & 1502.148 & 66571.34 & 0.006000 & & A \\*
    & 1703.412 & 58705.71 & 0.012240 &             & A & \smallcite{pickering00mnras} \\*
    & 1709.604 & 58493.07 & 0.032400 &   -20 (250) & A & \smallcite{pickering00mnras} \\*
    & 1741.553 & 57420.01 & 0.042700 & -1400 (250) & A & \smallcite{pickering00mnras} \\*
    & 1751.915 & 57080.37 & 0.027700 &  -700 (250) & A & \smallcite{pickering00mnras} \\*
Zn\textsc{II}
    & 2026.137 & 49355.00 & 0.489000 &  2479 (25) & M & \smallcite{aldenius06mnras,pickering00mnras}\smallcite{matsubara03apb} \\*
    & 2062.660 & 48481.08 & 0.256000 &  1584 (25) & I & \smallcite{aldenius06mnras,pickering00mnras} \\*
		\hline \hline
\end{longtable}

\bibliography{D:/references}

\begin{thebibliography}{34}
\expandafter\ifx\csname natexlab\endcsname\relax\def\natexlab#1{#1}\fi
\expandafter\ifx\csname bibnamefont\endcsname\relax
  \def\bibnamefont#1{#1}\fi
\expandafter\ifx\csname bibfnamefont\endcsname\relax
  \def\bibfnamefont#1{#1}\fi
\expandafter\ifx\csname citenamefont\endcsname\relax
  \def\citenamefont#1{#1}\fi
\expandafter\ifx\csname url\endcsname\relax
  \def\url#1{\texttt{#1}}\fi
\expandafter\ifx\csname urlprefix\endcsname\relax\def\urlprefix{URL }\fi
\providecommand{\bibinfo}[2]{#2}
\providecommand{\eprint}[2][]{\url{#2}}

\bibitem[{\citenamefont{Webb et~al.}(1999)\citenamefont{Webb, Flambaum,
  Churchill, Drinkwater, and Barrow}}]{webb99prl}
\bibinfo{author}{\bibfnamefont{J.~K.} \bibnamefont{Webb}},
  \bibinfo{author}{\bibfnamefont{V.~V.} \bibnamefont{Flambaum}},
  \bibinfo{author}{\bibfnamefont{C.~W.} \bibnamefont{Churchill}},
  \bibinfo{author}{\bibfnamefont{M.~J.} \bibnamefont{Drinkwater}},
  \bibnamefont{and} \bibinfo{author}{\bibfnamefont{J.~D.}
  \bibnamefont{Barrow}}, \bibinfo{journal}{\prl} \textbf{\bibinfo{volume}{82}},
  \bibinfo{pages}{884} (\bibinfo{year}{1999}).

\bibitem[{\citenamefont{Webb et~al.}(2001)\citenamefont{Webb, Murphy, Flambaum,
  Dzuba, Barrow, Churchill, Prochaska, and Wolfe}}]{webb01prl}
\bibinfo{author}{\bibfnamefont{J.~K.} \bibnamefont{Webb}},
  \bibinfo{author}{\bibfnamefont{M.~T.} \bibnamefont{Murphy}},
  \bibinfo{author}{\bibfnamefont{V.~V.} \bibnamefont{Flambaum}},
  \bibinfo{author}{\bibfnamefont{V.~A.} \bibnamefont{Dzuba}},
  \bibinfo{author}{\bibfnamefont{J.~D.} \bibnamefont{Barrow}},
  \bibinfo{author}{\bibfnamefont{C.~W.} \bibnamefont{Churchill}},
  \bibinfo{author}{\bibfnamefont{J.~X.} \bibnamefont{Prochaska}},
  \bibnamefont{and} \bibinfo{author}{\bibfnamefont{A.~M.} \bibnamefont{Wolfe}},
  \bibinfo{journal}{\prl} \textbf{\bibinfo{volume}{87}},
  \bibinfo{pages}{091301} (\bibinfo{year}{2001}).

\bibitem[{\citenamefont{Murphy et~al.}(2001{\natexlab{a}})\citenamefont{Murphy,
  Webb, Flambaum, Dzuba, Churchill, Prochaska, Barrow, and
  Wolfe}}]{murphy01mnrasA}
\bibinfo{author}{\bibfnamefont{M.~T.} \bibnamefont{Murphy}},
  \bibinfo{author}{\bibfnamefont{J.~K.} \bibnamefont{Webb}},
  \bibinfo{author}{\bibfnamefont{V.~V.} \bibnamefont{Flambaum}},
  \bibinfo{author}{\bibfnamefont{V.~A.} \bibnamefont{Dzuba}},
  \bibinfo{author}{\bibfnamefont{C.~W.} \bibnamefont{Churchill}},
  \bibinfo{author}{\bibfnamefont{J.~X.} \bibnamefont{Prochaska}},
  \bibinfo{author}{\bibfnamefont{J.~D.} \bibnamefont{Barrow}},
  \bibnamefont{and} \bibinfo{author}{\bibfnamefont{A.~M.} \bibnamefont{Wolfe}},
  \bibinfo{journal}{\mnras} \textbf{\bibinfo{volume}{327}},
  \bibinfo{pages}{1208} (\bibinfo{year}{2001}{\natexlab{a}}).

\bibitem[{\citenamefont{Murphy et~al.}(2001{\natexlab{b}})\citenamefont{Murphy,
  Webb, Flambaum, Prochaska, and Wolfe}}]{murphy01mnrasC}
\bibinfo{author}{\bibfnamefont{M.~T.} \bibnamefont{Murphy}},
  \bibinfo{author}{\bibfnamefont{J.~K.} \bibnamefont{Webb}},
  \bibinfo{author}{\bibfnamefont{V.~V.} \bibnamefont{Flambaum}},
  \bibinfo{author}{\bibfnamefont{J.~X.} \bibnamefont{Prochaska}},
  \bibnamefont{and} \bibinfo{author}{\bibfnamefont{A.~M.} \bibnamefont{Wolfe}},
  \bibinfo{journal}{\mnras} \textbf{\bibinfo{volume}{327}},
  \bibinfo{pages}{1237} (\bibinfo{year}{2001}{\natexlab{b}}).

\bibitem[{\citenamefont{Webb et~al.}(2003)\citenamefont{Webb, Murphy, Flambaum,
  and Curran}}]{webb03ass}
\bibinfo{author}{\bibfnamefont{J.~K.} \bibnamefont{Webb}},
  \bibinfo{author}{\bibfnamefont{M.~T.} \bibnamefont{Murphy}},
  \bibinfo{author}{\bibfnamefont{V.~V.} \bibnamefont{Flambaum}},
  \bibnamefont{and} \bibinfo{author}{\bibfnamefont{S.~J.}
  \bibnamefont{Curran}}, \bibinfo{journal}{\ass}
  \textbf{\bibinfo{volume}{283}}, \bibinfo{pages}{565} (\bibinfo{year}{2003}).

\bibitem[{\citenamefont{Murphy et~al.}(2003{\natexlab{a}})\citenamefont{Murphy,
  Webb, and Flambaum}}]{murphy03mnras}
\bibinfo{author}{\bibfnamefont{M.~T.} \bibnamefont{Murphy}},
  \bibinfo{author}{\bibfnamefont{J.~K.} \bibnamefont{Webb}}, \bibnamefont{and}
  \bibinfo{author}{\bibfnamefont{V.~V.} \bibnamefont{Flambaum}},
  \bibinfo{journal}{\mnras} \textbf{\bibinfo{volume}{345}},
  \bibinfo{pages}{609} (\bibinfo{year}{2003}{\natexlab{a}}).

\bibitem[{\citenamefont{Uzan}(2003)}]{uzan03rmp}
\bibinfo{author}{\bibfnamefont{J.-P.} \bibnamefont{Uzan}},
  \bibinfo{journal}{\rmp} \textbf{\bibinfo{volume}{75}}, \bibinfo{pages}{403}
  (\bibinfo{year}{2003}).

\bibitem[{\citenamefont{Quast et~al.}(2004)\citenamefont{Quast, Reimers, and
  Levshakov}}]{quast04aap}
\bibinfo{author}{\bibfnamefont{R.}~\bibnamefont{Quast}},
  \bibinfo{author}{\bibfnamefont{D.}~\bibnamefont{Reimers}}, \bibnamefont{and}
  \bibinfo{author}{\bibfnamefont{S.~A.} \bibnamefont{Levshakov}},
  \bibinfo{journal}{\aap} \textbf{\bibinfo{volume}{414}}, \bibinfo{pages}{L7}
  (\bibinfo{year}{2004}).

\bibitem[{\citenamefont{Srianand et~al.}(2004)\citenamefont{Srianand, Chand,
  Petitjean, and Aracil}}]{srianand04prl}
\bibinfo{author}{\bibfnamefont{R.}~\bibnamefont{Srianand}},
  \bibinfo{author}{\bibfnamefont{H.}~\bibnamefont{Chand}},
  \bibinfo{author}{\bibfnamefont{P.}~\bibnamefont{Petitjean}},
  \bibnamefont{and} \bibinfo{author}{\bibfnamefont{B.}~\bibnamefont{Aracil}},
  \bibinfo{journal}{\prl} \textbf{\bibinfo{volume}{92}},
  \bibinfo{pages}{121302} (\bibinfo{year}{2004}).

\bibitem[{\citenamefont{Chand et~al.}(2004)\citenamefont{Chand, Srianand,
  Petitjean, and Aracil}}]{chand04aap}
\bibinfo{author}{\bibfnamefont{H.}~\bibnamefont{Chand}},
  \bibinfo{author}{\bibfnamefont{R.}~\bibnamefont{Srianand}},
  \bibinfo{author}{\bibfnamefont{P.}~\bibnamefont{Petitjean}},
  \bibnamefont{and} \bibinfo{author}{\bibfnamefont{B.}~\bibnamefont{Aracil}},
  \bibinfo{journal}{\aap} \textbf{\bibinfo{volume}{417}}, \bibinfo{pages}{853}
  (\bibinfo{year}{2004}).

\bibitem[{\citenamefont{Levshakov et~al.}(2005)\citenamefont{Levshakov,
  Centuri\'on, Molaro, and D'Odorico}}]{levshakov05aap}
\bibinfo{author}{\bibfnamefont{S.~A.} \bibnamefont{Levshakov}},
  \bibinfo{author}{\bibfnamefont{M.}~\bibnamefont{Centuri\'on}},
  \bibinfo{author}{\bibfnamefont{P.}~\bibnamefont{Molaro}}, \bibnamefont{and}
  \bibinfo{author}{\bibfnamefont{S.}~\bibnamefont{D'Odorico}},
  \bibinfo{journal}{\aap} \textbf{\bibinfo{volume}{434}}, \bibinfo{pages}{827}
  (\bibinfo{year}{2005}).

\bibitem[{\citenamefont{Morton}(1991)}]{morton91ajss}
\bibinfo{author}{\bibfnamefont{D.~C.} \bibnamefont{Morton}},
  \bibinfo{journal}{\ajss} \textbf{\bibinfo{volume}{77}}, \bibinfo{pages}{119}
  (\bibinfo{year}{1991}).

\bibitem[{\citenamefont{Morton}(2003)}]{morton03ajss}
\bibinfo{author}{\bibfnamefont{D.~C.} \bibnamefont{Morton}},
  \bibinfo{journal}{\ajss} \textbf{\bibinfo{volume}{149}}, \bibinfo{pages}{205}
  (\bibinfo{year}{2003}).

\bibitem[{\citenamefont{Murphy et~al.}(2001{\natexlab{c}})\citenamefont{Murphy,
  Webb, Flambaum, Churchill, and Prochaska}}]{murphy01mnrasB}
\bibinfo{author}{\bibfnamefont{M.~T.} \bibnamefont{Murphy}},
  \bibinfo{author}{\bibfnamefont{J.~K.} \bibnamefont{Webb}},
  \bibinfo{author}{\bibfnamefont{V.~V.} \bibnamefont{Flambaum}},
  \bibinfo{author}{\bibfnamefont{C.~W.} \bibnamefont{Churchill}},
  \bibnamefont{and} \bibinfo{author}{\bibfnamefont{J.~X.}
  \bibnamefont{Prochaska}}, \bibinfo{journal}{\mnras}
  \textbf{\bibinfo{volume}{327}}, \bibinfo{pages}{1223}
  (\bibinfo{year}{2001}{\natexlab{c}}).

\bibitem[{\citenamefont{Murphy et~al.}(2003{\natexlab{b}})\citenamefont{Murphy,
  Webb, Flambaum, and Curran}}]{murphy03ass}
\bibinfo{author}{\bibfnamefont{M.~T.} \bibnamefont{Murphy}},
  \bibinfo{author}{\bibfnamefont{J.~K.} \bibnamefont{Webb}},
  \bibinfo{author}{\bibfnamefont{V.~V.} \bibnamefont{Flambaum}},
  \bibnamefont{and} \bibinfo{author}{\bibfnamefont{S.~J.}
  \bibnamefont{Curran}}, \bibinfo{journal}{\ass}
  \textbf{\bibinfo{volume}{283}}, \bibinfo{pages}{577}
  (\bibinfo{year}{2003}{\natexlab{b}}).

\bibitem[{\citenamefont{Dzuba et~al.}(2002)\citenamefont{Dzuba, Flambaum,
  Kozlov, and Marchenko}}]{dzuba02praA}
\bibinfo{author}{\bibfnamefont{V.~A.} \bibnamefont{Dzuba}},
  \bibinfo{author}{\bibfnamefont{V.~V.} \bibnamefont{Flambaum}},
  \bibinfo{author}{\bibfnamefont{M.~G.} \bibnamefont{Kozlov}},
  \bibnamefont{and}
  \bibinfo{author}{\bibfnamefont{M.}~\bibnamefont{Marchenko}},
  \bibinfo{journal}{\pra} \textbf{\bibinfo{volume}{66}},
  \bibinfo{pages}{022501} (\bibinfo{year}{2002}).

\bibitem[{\citenamefont{Berengut et~al.}(2004)\citenamefont{Berengut, Dzuba,
  Flambaum, and Marchenko}}]{berengut04praB}
\bibinfo{author}{\bibfnamefont{J.~C.} \bibnamefont{Berengut}},
  \bibinfo{author}{\bibfnamefont{V.~A.} \bibnamefont{Dzuba}},
  \bibinfo{author}{\bibfnamefont{V.~V.} \bibnamefont{Flambaum}},
  \bibnamefont{and} \bibinfo{author}{\bibfnamefont{M.~V.}
  \bibnamefont{Marchenko}}, \bibinfo{journal}{\pra}
  \textbf{\bibinfo{volume}{70}}, \bibinfo{pages}{064101}
  (\bibinfo{year}{2004}).

\bibitem[{\citenamefont{Berengut et~al.}(2005)\citenamefont{Berengut, Flambaum,
  and Kozlov}}]{berengut05pra}
\bibinfo{author}{\bibfnamefont{J.~C.} \bibnamefont{Berengut}},
  \bibinfo{author}{\bibfnamefont{V.~V.} \bibnamefont{Flambaum}},
  \bibnamefont{and} \bibinfo{author}{\bibfnamefont{M.~G.}
  \bibnamefont{Kozlov}}, \bibinfo{journal}{\pra} \textbf{\bibinfo{volume}{72}},
  \bibinfo{pages}{044501} (\bibinfo{year}{2005}).

\bibitem[{\citenamefont{Berengut et~al.}(2006)\citenamefont{Berengut, Flambaum,
  and Kozlov}}]{berengut06pra}
\bibinfo{author}{\bibfnamefont{J.~C.} \bibnamefont{Berengut}},
  \bibinfo{author}{\bibfnamefont{V.~V.} \bibnamefont{Flambaum}},
  \bibnamefont{and} \bibinfo{author}{\bibfnamefont{M.~G.}
  \bibnamefont{Kozlov}}, \bibinfo{journal}{\pra} \textbf{\bibinfo{volume}{73}},
  \bibinfo{pages}{012504} (\bibinfo{year}{2006}).

\bibitem[{\citenamefont{Labazan et~al.}(2005)\citenamefont{Labazan, Reinhold,
  Ubachs, and Flambaum}}]{labazan05pra}
\bibinfo{author}{\bibfnamefont{I.}~\bibnamefont{Labazan}},
  \bibinfo{author}{\bibfnamefont{E.}~\bibnamefont{Reinhold}},
  \bibinfo{author}{\bibfnamefont{W.}~\bibnamefont{Ubachs}}, \bibnamefont{and}
  \bibinfo{author}{\bibfnamefont{V.~V.} \bibnamefont{Flambaum}},
  \bibinfo{journal}{\pra} \textbf{\bibinfo{volume}{71}},
  \bibinfo{pages}{040501} (\bibinfo{year}{2005}).

\bibitem[{\citenamefont{Griesmann and Kling}(2000)}]{griesmann00apj}
\bibinfo{author}{\bibfnamefont{U.}~\bibnamefont{Griesmann}} \bibnamefont{and}
  \bibinfo{author}{\bibfnamefont{R.}~\bibnamefont{Kling}},
  \bibinfo{journal}{\apj} \textbf{\bibinfo{volume}{536}}, \bibinfo{pages}{L113}
  (\bibinfo{year}{2000}).

\bibitem[{\citenamefont{Juncar et~al.}(1981)\citenamefont{Juncar, Pinard,
  Hamon, and Chartier}}]{juncar81met}
\bibinfo{author}{\bibfnamefont{P.}~\bibnamefont{Juncar}},
  \bibinfo{author}{\bibfnamefont{J.}~\bibnamefont{Pinard}},
  \bibinfo{author}{\bibfnamefont{J.}~\bibnamefont{Hamon}}, \bibnamefont{and}
  \bibinfo{author}{\bibfnamefont{A.}~\bibnamefont{Chartier}},
  \bibinfo{journal}{\met} \textbf{\bibinfo{volume}{17}}, \bibinfo{pages}{77}
  (\bibinfo{year}{1981}).

\bibitem[{\citenamefont{Pescht et~al.}(1977)\citenamefont{Pescht, Gerhardt, and
  Matthias}}]{pescht77zpa}
\bibinfo{author}{\bibfnamefont{K.}~\bibnamefont{Pescht}},
  \bibinfo{author}{\bibfnamefont{H.}~\bibnamefont{Gerhardt}}, \bibnamefont{and}
  \bibinfo{author}{\bibfnamefont{E.}~\bibnamefont{Matthias}},
  \bibinfo{journal}{\zpa} \textbf{\bibinfo{volume}{281}}, \bibinfo{pages}{199}
  (\bibinfo{year}{1977}).

\bibitem[{\citenamefont{Huber et~al.}(1978)\citenamefont{Huber, Touchard,
  Büttgenbach, Thibault, Klapisch, Duong, Liberman, Pinard, Vialle, Juncar
  et~al.}}]{huber78prc}
\bibinfo{author}{\bibfnamefont{G.}~\bibnamefont{Huber}},
  \bibinfo{author}{\bibfnamefont{F.}~\bibnamefont{Touchard}},
  \bibinfo{author}{\bibfnamefont{S.}~\bibnamefont{Büttgenbach}},
  \bibinfo{author}{\bibfnamefont{C.}~\bibnamefont{Thibault}},
  \bibinfo{author}{\bibfnamefont{R.}~\bibnamefont{Klapisch}},
  \bibinfo{author}{\bibfnamefont{H.~T.} \bibnamefont{Duong}},
  \bibinfo{author}{\bibfnamefont{S.}~\bibnamefont{Liberman}},
  \bibinfo{author}{\bibfnamefont{J.}~\bibnamefont{Pinard}},
  \bibinfo{author}{\bibfnamefont{J.~L.} \bibnamefont{Vialle}},
  \bibinfo{author}{\bibfnamefont{P.}~\bibnamefont{Juncar}},
  \bibnamefont{et~al.}, \bibinfo{journal}{\prc} \textbf{\bibinfo{volume}{18}},
  \bibinfo{pages}{2342} (\bibinfo{year}{1978}).

\bibitem[{\citenamefont{Gangrsky et~al.}(1998)\citenamefont{Gangrsky,
  Karaivanov, Marinova, Markov, Melnikova, Mishinsky, Zemlyanoi, and
  Zhemenik}}]{gangrsky98epja}
\bibinfo{author}{\bibfnamefont{Y.~P.} \bibnamefont{Gangrsky}},
  \bibinfo{author}{\bibfnamefont{D.~V.} \bibnamefont{Karaivanov}},
  \bibinfo{author}{\bibfnamefont{K.~P.} \bibnamefont{Marinova}},
  \bibinfo{author}{\bibfnamefont{B.~N.} \bibnamefont{Markov}},
  \bibinfo{author}{\bibfnamefont{L.~M.} \bibnamefont{Melnikova}},
  \bibinfo{author}{\bibfnamefont{G.~V.} \bibnamefont{Mishinsky}},
  \bibinfo{author}{\bibfnamefont{S.~G.} \bibnamefont{Zemlyanoi}},
  \bibnamefont{and} \bibinfo{author}{\bibfnamefont{V.~I.}
  \bibnamefont{Zhemenik}}, \bibinfo{journal}{\epja}
  \textbf{\bibinfo{volume}{3}}, \bibinfo{pages}{313} (\bibinfo{year}{1998}).

\bibitem[{\citenamefont{Aldenius et~al.}(2006)\citenamefont{Aldenius,
  Johansson, and Murphy}}]{aldenius06mnras}
\bibinfo{author}{\bibfnamefont{M.}~\bibnamefont{Aldenius}},
  \bibinfo{author}{\bibfnamefont{S.}~\bibnamefont{Johansson}},
  \bibnamefont{and} \bibinfo{author}{\bibfnamefont{M.~T.}
  \bibnamefont{Murphy}}, \bibinfo{journal}{submitted to \mnras}
  (\bibinfo{year}{2006}).

\bibitem[{\citenamefont{Pickering et~al.}(1998)\citenamefont{Pickering, Thorne,
  and Webb}}]{pickering98mnras}
\bibinfo{author}{\bibfnamefont{J.~C.} \bibnamefont{Pickering}},
  \bibinfo{author}{\bibfnamefont{A.~P.} \bibnamefont{Thorne}},
  \bibnamefont{and} \bibinfo{author}{\bibfnamefont{J.~K.} \bibnamefont{Webb}},
  \bibinfo{journal}{\mnras} \textbf{\bibinfo{volume}{300}},
  \bibinfo{pages}{131–134} (\bibinfo{year}{1998}).

\bibitem[{\citenamefont{Hallstadius}(1979)}]{hallstadius79zpa}
\bibinfo{author}{\bibfnamefont{L.}~\bibnamefont{Hallstadius}},
  \bibinfo{journal}{\zpa} \textbf{\bibinfo{volume}{291}}, \bibinfo{pages}{203}
  (\bibinfo{year}{1979}).

\bibitem[{\citenamefont{Boiteux et~al.}(1988)\citenamefont{Boiteux, Klein,
  Leite, and Ducloy}}]{boiteux88jpf}
\bibinfo{author}{\bibfnamefont{S.~L.} \bibnamefont{Boiteux}},
  \bibinfo{author}{\bibfnamefont{A.}~\bibnamefont{Klein}},
  \bibinfo{author}{\bibfnamefont{J.~R.~R.} \bibnamefont{Leite}},
  \bibnamefont{and} \bibinfo{author}{\bibfnamefont{M.}~\bibnamefont{Ducloy}},
  \bibinfo{journal}{\jpf} \textbf{\bibinfo{volume}{49}}, \bibinfo{pages}{885}
  (\bibinfo{year}{1988}).

\bibitem[{\citenamefont{Drullinger et~al.}(1980)\citenamefont{Drullinger,
  Wineland, and Bergquist}}]{drullinger80ap}
\bibinfo{author}{\bibfnamefont{R.~E.} \bibnamefont{Drullinger}},
  \bibinfo{author}{\bibfnamefont{D.}~\bibnamefont{Wineland}}, \bibnamefont{and}
  \bibinfo{author}{\bibfnamefont{J.~C.} \bibnamefont{Bergquist}},
  \bibinfo{journal}{\ap} \textbf{\bibinfo{volume}{22}}, \bibinfo{pages}{365}
  (\bibinfo{year}{1980}).

\bibitem[{\citenamefont{Pickering et~al.}(2000)\citenamefont{Pickering, Thorne,
  Murray, Litz\'en, Johansson, Zilio, and Webb}}]{pickering00mnras}
\bibinfo{author}{\bibfnamefont{J.~C.} \bibnamefont{Pickering}},
  \bibinfo{author}{\bibfnamefont{A.~P.} \bibnamefont{Thorne}},
  \bibinfo{author}{\bibfnamefont{J.~E.} \bibnamefont{Murray}},
  \bibinfo{author}{\bibfnamefont{U.}~\bibnamefont{Litz\'en}},
  \bibinfo{author}{\bibfnamefont{S.}~\bibnamefont{Johansson}},
  \bibinfo{author}{\bibfnamefont{V.}~\bibnamefont{Zilio}}, \bibnamefont{and}
  \bibinfo{author}{\bibfnamefont{J.~K.} \bibnamefont{Webb}},
  \bibinfo{journal}{\mnras} \textbf{\bibinfo{volume}{319}},
  \bibinfo{pages}{163} (\bibinfo{year}{2000}).

\bibitem[{\citenamefont{Pickering et~al.}(2002)\citenamefont{Pickering,
  Donnelly, Nilsson, Hibbert, and Johansson}}]{pickering02aap}
\bibinfo{author}{\bibfnamefont{J.~C.} \bibnamefont{Pickering}},
  \bibinfo{author}{\bibfnamefont{M.~P.} \bibnamefont{Donnelly}},
  \bibinfo{author}{\bibfnamefont{H.}~\bibnamefont{Nilsson}},
  \bibinfo{author}{\bibfnamefont{A.}~\bibnamefont{Hibbert}}, \bibnamefont{and}
  \bibinfo{author}{\bibfnamefont{S.}~\bibnamefont{Johansson}},
  \bibinfo{journal}{\aap} \textbf{\bibinfo{volume}{396}},
  \bibinfo{pages}{715–722} (\bibinfo{year}{2002}).

\bibitem[{\citenamefont{Nave et~al.}(1991)\citenamefont{Nave, Learner, Thorne,
  and Harris}}]{nave91josab}
\bibinfo{author}{\bibfnamefont{G.}~\bibnamefont{Nave}},
  \bibinfo{author}{\bibfnamefont{R.~C.~M.} \bibnamefont{Learner}},
  \bibinfo{author}{\bibfnamefont{A.~P.} \bibnamefont{Thorne}},
  \bibnamefont{and} \bibinfo{author}{\bibfnamefont{C.~J.}
  \bibnamefont{Harris}}, \bibinfo{journal}{\josab}
  \textbf{\bibinfo{volume}{8}}, \bibinfo{pages}{2028} (\bibinfo{year}{1991}).

\bibitem[{\citenamefont{Matsubara et~al.}(2002)\citenamefont{Matsubara, Tanaka,
  Imajo, Urabe, and Watanabe}}]{matsubara03apb}
\bibinfo{author}{\bibfnamefont{K.}~\bibnamefont{Matsubara}},
  \bibinfo{author}{\bibfnamefont{U.}~\bibnamefont{Tanaka}},
  \bibinfo{author}{\bibfnamefont{H.}~\bibnamefont{Imajo}},
  \bibinfo{author}{\bibfnamefont{S.}~\bibnamefont{Urabe}}, \bibnamefont{and}
  \bibinfo{author}{\bibfnamefont{M.}~\bibnamefont{Watanabe}},
  \bibinfo{journal}{\apb} \textbf{\bibinfo{volume}{76}}, \bibinfo{pages}{209}
  (\bibinfo{year}{2002}).

\end{thebibliography}

\end{document}